\documentclass[twocolumn]{aastex63}
\usepackage{color}

\newcommand{\black}{\color{black}}
\newcommand{\red}{\black}

\received{May 15}
\revised{July 3}
\accepted{July 15, 2021}

\submitjournal{ApJL}

\shorttitle{300-TeV photons associated with a neutrino}
\shortauthors{Carpet--3 Group}
\graphicspath{{./}{figures/}}

\begin{document}


\title{Observation of photons above 300~TeV associated with a high-energy
neutrino from the Cygnus region}

\correspondingauthor{Sergey Troitsky}
\email{st@ms2.inr.ac.ru}

\author[0000-0001-9956-5439]{D.~D.~Dzhappuev}
\affiliation{Institute for Nuclear Research of the Russian Academy of Sciences, 60th October Anniversary Prospect 7a, Moscow 117312, Russia}
\author[0000-0003-2652-0293]{Yu.~Z.~Afashokov}
\affiliation{Institute for Nuclear Research of the Russian Academy of Sciences, 60th October Anniversary Prospect 7a, Moscow 117312, Russia}
\author{I.~M.~Dzaparova}
\affiliation{Institute for Nuclear Research of the Russian Academy of Sciences, 60th October Anniversary Prospect 7a, Moscow 117312, Russia}
\affiliation{Institute of Astronomy, Russian Academy of Sciences, 119017, Moscow, Russia}
\author[0000-0002-7660-4236]{T.~A.~Dzhatdoev}
\affiliation{D.V.~Skobeltsyn Institute of Nuclear Physics, M. V. Lomonosov Moscow State University, Moscow 119991, Russia}
\affiliation{Institute for Nuclear Research of the Russian Academy of Sciences, 60th October Anniversary Prospect 7a, Moscow 117312, Russia}
\author{E.~A.~Gorbacheva}
\affiliation{Institute for Nuclear Research of the Russian Academy of Sciences, 60th October Anniversary Prospect 7a, Moscow 117312, Russia}
\author{I.~S.~Karpikov}
\affiliation{Institute for Nuclear Research of the Russian Academy of Sciences, 60th October Anniversary Prospect 7a, Moscow 117312, Russia}
\author[0000-0002-0564-8477]{M.~M.~Khadzhiev}
\affiliation{Institute for Nuclear Research of the Russian Academy of Sciences, 60th October Anniversary Prospect 7a, Moscow 117312, Russia}
\author[0000-0003-4301-9599]{N.~F.~Klimenko}
\affiliation{Institute for Nuclear Research of the Russian Academy of Sciences, 60th October Anniversary Prospect 7a, Moscow 117312, Russia}
\author[0000-0002-8337-3878]{A.~U.~Kudzhaev}
\affiliation{Institute for Nuclear Research of the Russian Academy of Sciences, 60th October Anniversary Prospect 7a, Moscow 117312, Russia}
\author{A.~N.~Kurenya}
\affiliation{Institute for Nuclear Research of the Russian Academy of Sciences, 60th October Anniversary Prospect 7a, Moscow 117312, Russia}
\author[0000-0002-8002-6491]{A.~S.~Lidvansky}
\affiliation{Institute for Nuclear Research of the Russian Academy of Sciences, 60th October Anniversary Prospect 7a, Moscow 117312, Russia}
\author[0000-0001-6146-8318]{O.~I.~Mikhailova}
\affiliation{Institute for Nuclear Research of the Russian Academy of Sciences, 60th October Anniversary Prospect 7a, Moscow 117312, Russia}
\author{V.~B.~Petkov}
\affiliation{Institute for Nuclear Research of the Russian Academy of Sciences, 60th October Anniversary Prospect 7a, Moscow 117312, Russia}
\affiliation{Institute of Astronomy, Russian Academy of Sciences, 119017, Moscow, Russia}
\author[0000-0003-3395-0419]{E.~I.~Podlesnyi}
\affiliation{Physics Department, M. V. Lomonosov Moscow State University, Moscow 119991, Russia}
\affiliation{D.V.~Skobeltsyn Institute of Nuclear Physics, M. V. Lomonosov Moscow State University, Moscow 119991, Russia}
\affiliation{Institute for Nuclear Research of the Russian Academy of Sciences, 60th October Anniversary Prospect 7a, Moscow 117312, Russia}
\author[0000-0001-9784-2244]{V.~S.~Romanenko}
\affiliation{Institute for Nuclear Research of the Russian Academy of Sciences, 60th October Anniversary Prospect 7a, Moscow 117312, Russia}
\author[0000-0002-6106-2673]{G.~I.~Rubtsov}
\affiliation{Institute for Nuclear Research of the Russian Academy of Sciences, 60th October Anniversary Prospect 7a, Moscow 117312, Russia}
\author[0000-0001-6917-6600]{S.~V.~Troitsky}
\affiliation{Institute for Nuclear Research of the Russian Academy of Sciences, 60th October Anniversary Prospect 7a, Moscow 117312, Russia}
\author{I.~B.~Unatlokov}
\affiliation{Institute for Nuclear Research of the Russian Academy of Sciences, 60th October Anniversary Prospect 7a, Moscow 117312, Russia}
\author[0000-0002-8255-3631]{I.~A.~Vaiman}
\affiliation{Physics Department, M. V. Lomonosov Moscow State University, Moscow 119991, Russia}
\affiliation{D.V.~Skobeltsyn Institute of Nuclear Physics, M. V. Lomonosov Moscow State University, Moscow 119991, Russia}
\author{A.~F.~Yanin}
\affiliation{Institute for Nuclear Research of the Russian Academy of Sciences, 60th October Anniversary Prospect 7a, Moscow 117312, Russia}
\author[0000-0002-3588-9706]{Ya.~V.~Zhezher}
\affiliation{Institute for Nuclear Research of the Russian Academy of Sciences, 60th October Anniversary Prospect 7a, Moscow 117312, Russia}
\affiliation{Institute for Cosmic Ray Research, University of Tokyo, Kashiwa, Kashiwanoha, 5-1-5, 277-8582, Japan}
\author[0000-0001-6086-4247]{K.~V.~Zhuravleva}
\affiliation{Institute for Nuclear Research of the Russian Academy of Sciences, 60th October Anniversary Prospect 7a, Moscow 117312, Russia}
\collaboration{22}{(Carpet--3 Group)}



\begin{abstract}
Galactic sites of acceleration of cosmic rays to energies of order $10^{15}$~eV and higher, dubbed PeVatrons, reveal themselves by recently discovered gamma radiation of energies above 100~TeV. However,  joint gamma-ray and neutrino production, which marks unambiguously cosmic-ray interactions with ambient matter and radiation, was not observed until now. In November 2020, the IceCube neutrino observatory reported an $\sim150$ TeV neutrino event from the direction of one of the most promising Galactic PeVatrons, the Cygnus Cocoon. Here we report on the observation of a 3.1-sigma (post trial) excess of atmospheric air showers from the same direction, observed by the Carpet-2 experiment and consistent with a few-months flare in photons above 300~TeV, in temporal coincidence with the neutrino event. The fluence of the gamma-ray flare is of the same order as that expected from the neutrino observation, assuming the standard mechanism of neutrino production. This is the first evidence for the joint production of high-energy neutrinos and gamma rays in a Galactic source.
\end{abstract}



\section{Introduction} 
\label{sec:intro}
Recent observations of astrophysical gamma rays above 100~TeV established the existence of various Galactic sources, both point-like \citep{HAWC-Crab-100TeV,Tibet-Crab-100TeV,HAWC56-100,HAWC-G106,HAWC-J1825-200TeV,LHAASO-12sources} and diffuse \citep{Tibet-GalDiffuse}. These observations are often interpreted as indications to the existence of Galactic PeVatrons, that are sites of cosmic-ray acceleration up to $\gtrsim$~PeV energies\footnote{1~PeV$=10^{15}$~eV.}, in which the gamma rays are produced in interactions of energetic hadrons with ambient matter and radiation. Observations of neutrinos co-produced with these gamma rays would unambiguously point to their hadronic origin.


\red It is an intriguing question whether some of the high-energy (above TeV) astrophysical neutrinos (\citealt{IceCubeFirstPeV,IceCubeFirst26}; for recent reviews, see e.g. \citealt{AhlersHalzen,VissaniUniverse}) come from Galactic sources. While the largest available data set of the IceCube and ANTARES experiments does not demonstrate any correlation of neutrino arrival directions with the Galactic disk \citep{IceCubeANTARES-Gal}, various indications exist in favor of \black the Galactic origin of a part of the neutrino flux at energies below $\sim200$~TeV. The most recent analysis of arrival directions of IceCube cascade events \citep{IceCube-CascadeMap-Gal} reveals a weak Galactic-plane excess. Studies of track-like and cascade-like events registered in the IceCube detector under the assumption of the power-law shape of the primary neutrino spectrum yield \citep[e.g.,][]{IceCube2020-HESE} different values of the power-law index for these two samples of events. This discrepancy may be naturally explained if the primary spectrum is actually composed of two distinct components \citep{2comp-Chen}. A population of extragalactic sources which demonstrates significant correlation with astrophysical neutrinos \citep{neutradio-low} may be responsible for the hard component, while the soft component may be due to the Galactic sources \citep{2comp-Vissani,2comp-Vissani-2}. In the latter case, the extragalactic gamma-ray background (EGRB) component concomitant with IceCube neutrinos does not overshoot the EGRB measured with Fermi-LAT.


Cygnus Cocoon \citep{Fermi-Cocoon2011}, an extended gamma-ray source presumably containing an OB star association embedded in a superbubble, \red is a candidate Galactic hadronic PeVatron. Star-forming regions \black are potential sites of cosmic-ray acceleration, gamma-ray and neutrino production \citep{Bykov-reviewHE}. In particular, it has been predicted that the flux of high-energy neutrinos from Cygnus Cocoon is close to the IceCube sensitivity \citep{Yoast-Hull-Cocoon-nu}. This source was detected by HAWC up to, and possibly beyond 100~TeV \citep{HAWC56-100,HAWC-Cocoon}\red; its position is consistent with the highest-energy (up to 1.4~PeV) photon source detected by LHAASO \citep{LHAASO-12sources}. Gamma-ray sources \black in the Cygnus region contribute a lot to the Galactic-plane diffuse gamma radiation above 400~TeV, discovered by \citet{Tibet-GalDiffuse}.

On November 20, 2020, IceCube reported \citep{IceCube-GCN.28927} a candidate track-like neutrino event with \red the estimated energy of 154~TeV. \black The arrival direction of the neutrino, though determined with a considerably low precision, coincided with the direction from Cygnus Cocoon. The event was reported through the standard BRONZE alert procedure \citep{IceCube-Gold-Bronze-alerts}. These alerts are routinely followed up by numerous instruments, in particular \citep{CarpetJETPLalerts2020} by the Carpet--2 gamma-ray telescope at the Baksan Neutrino Observatory. This event, however, was exceptional in the sense that it coincided with a previously defined prospective Galactic source of high-energy neutrinos. This gives a chance to detect sub-PeV gamma rays co-produced with neutrinos, which cannot reach us from extragalactic sources because of pair production on cosmic microwave background radiation \citep{Nikishov1962}. Standard Carpet--2 alert analysis revealed two gamma-ray candidate events in one-month interval centered at the alert time \citep{CarpetATel-obs}. Here, we present results of a more detailed study of a possible sub-PeV gamma-ray flare in the Cygnus Cocoon associated with the IceCube neutrino event. 

\section{The Carpet--2 detector and the dataset}
\label{sec:Carpet2}
Carpet--2 is an air-shower experiment co-located with the Baksan Neutrino Observatory (Neutrino village, North Caucasus). It includes a 200~m$^2$ continuous central scintillator detector, Carpet; five outer detector stations with 9~m$^2$ of scintillator in each of them; and a 175~m$^2$ shielded detector  with the threshold of 1~GeV for vertical muons. The primary-particle energy is determined from the shower size $N_e$, reconstructed from the central Carpet; the arrival direction is obtained from timing of the outer stations; the muon detector is used to select candidate gamma-ray showers which are muon-poor. The installation, its operation and data processing are described by \citet{Carpet2-description2007,Carpet2-Szabelski2009,Carpet2-arXiv2015,Carpet2019ICRC...36..808T,Carpet2019TeVPA-point}.  

The angular resolution of Carpet--2 is determined by a combination of (i)~fluctuations in the shower, (ii)~fluctuations in electronics and (iii)~earlier trigger of individual detector station due to coincident atmospheric muons. The point-spread function (PSF) has been determined experimentally \citep{Carpet2-CherenkovAngResPreprint} by means of simultaneous observations of air showers by Carpet and by an atmospheric Cerenkov detector (CD). The pointing accuracy of the CD, 0.1$^\circ$, and its angular resolution, 0.6$^\circ$, have been \red measured \black from observations of bright stars. The PSF of Carpet is non-Gaussian; 86\% of events are reconstructed within 4.7$^\circ$ of their true direction. Monte-Carlo simulations and experimental measurements of individual contributions (i), (ii), (iii) give results consistent with this estimate.

For the present study, we use Carpet--2 data recorded between April 7, 2018 and April 25, 2021, total 829 days of data \red collection\black. Standard quality cuts require that $\ge500$~GeV air-shower energy is deposited in Carpet; four outer stations participate in the determination of the arrival direction; the reconstructed shower axis is at least 0.7~m within the Carpet boundary; the reconstructed zenith angle is $\le 40^\circ$. In total, 65703 events passed these cuts in this time period.

\section{Simulations and data analysis}

Monte-Carlo simulations we use include air-shower modelling and the detector-response simulation described by \citet{Carpet2019JETPLneutrino-old}. Every simulated event is recorded in the same format and reconstructed by the same program as those used for the real data, including the quality-cut selection. These simulations are used to relate the reconstructed shower size $N_e$ to the primary gamma-ray energy $E_\gamma$, to estimate the detection and reconstruction efficiency and to develop criteria for separation between events induced by primary photons and by cosmic rays. Since the efficiency of the detection of gamma-ray events drops fastly below $E_\gamma\sim300$~TeV \citep{Carpet2019ICRC...36..808T}, we include only 56969 events with reconstructed $E_\gamma>300$~TeV in the data sample we use here. The effective area of the installation as the function of energy is presented by \citet{CarpetJETPLalerts2020}. Carpet--2 tests the same range of \red energies and \black fluxes for gamma rays as IceCube tests for neutrinos.

We also determine the notion of a ``photon median candidate event'' \citep{AugerSDgamma2007} from simulations as follows. Assuming the $E_\gamma^{-2}$ primary spectrum, we simulate a large number of gamma-ray induced events and select those with reconstructed $E_\gamma>300$~TeV. For each of those events, we calculate the ratio of the number of muons in the shielded detector, $n_\mu$, to $N_e$ and select the value $\alpha$ such that 50\% of reconstructed gamma rays satisfy $n_\mu/N_e<\alpha$. In the search for gamma-ray flares of localized sources, when the isotropic and uniform in time cosmic-ray background is small, it makes sense to use also the directional and temporal coincidence as a distinctive criterion for primary photons: cosmic-ray particles are charged and, at the sub-PeV energies we study here, isotropized in their directions and smeared in arrival times by the Galactic magnetic field. In this work, we use the entire sample to search for the gamma-ray excess associated with the neutrino event, then repeat the same procedure for the events selected by the ``photon median'' cut and check that the results are consistent. In this way, we both increase the available statistics and make the study less sensitive to the assumption about the source gamma-ray spectrum.

Previous Monte-Carlo simulations \red for this \citep{Carpet2-CherenkovAngResPreprint} and other \citet{BLLpredicitons} air-shower experiments \black indicated that counting of events within \red the (86--90)\% \black PSF containment angle is optimal in terms of the signal-to-noise ratio for point-source searches. In what follows, we concentrate on the circular region in the sky of this \red (86\% PSF) \black 4.7$^\circ$ angular radius, centered at the 4th Fermi-LAT source catalog 4FGL \citep{4FGL} best-fit position of the source \object{4FGL~J2028.6$+$4110e}, associated with the Cygnus Cocoon, and call this region the Cygnus-Cocoon Circle (CCC).

In the full 3-year Carpet--2 data sets, defined above, the number of events in CCC is consistent with that expected from random background, so the source is not detected significantly above 300~TeV. We obtain an upper limit on its integral gamma-ray flux as $I_\gamma(E_\gamma>300~\mbox{TeV})<2.6 \times 10^{-13}$~cm$^{-2}$s$^{-1}$ (95\% CL). The situation is different when the time period around the neutrino event is considered.

To search for a flux enhancement around the neutrino event, we proceed as follows. Denote the total number of events observed from the CCC as $N$. Since we have no prejudice about the possible \red flare \black time window, we consider the duration of the putative flare as a free parameter $d$ and vary $d/2$ between 1 and 60 days around the neutrino event, in steps of 1 day. For each of these 60 time windows, we determine the number $M(d)$ of events from CCC in this period and calculate the binomial probability $p(d)$ to observe this or larger number of events assuming constant mean rate of events per day. We then find the pre-trial p-value $p_{\rm pre}=\min\limits_d p(d)$. To estimate the effect of multiple trials, we perform a Monte-Carlo simulation of $10^4$ mock sets of arrival times of $N$ events. For each mock set $i$, we repeat the same procedure of varying the window width and finding the mock pre-trial p-value, $p_i$. The fraction of mock sets with $p_i \le p_{\rm pre}$ gives the post-trial probability $p$ which determines the significance of the observed effect, if any.

\section{Results and discussion}
\label{sec:results}
\begin{figure}
    \centering
    \includegraphics[width=\linewidth]{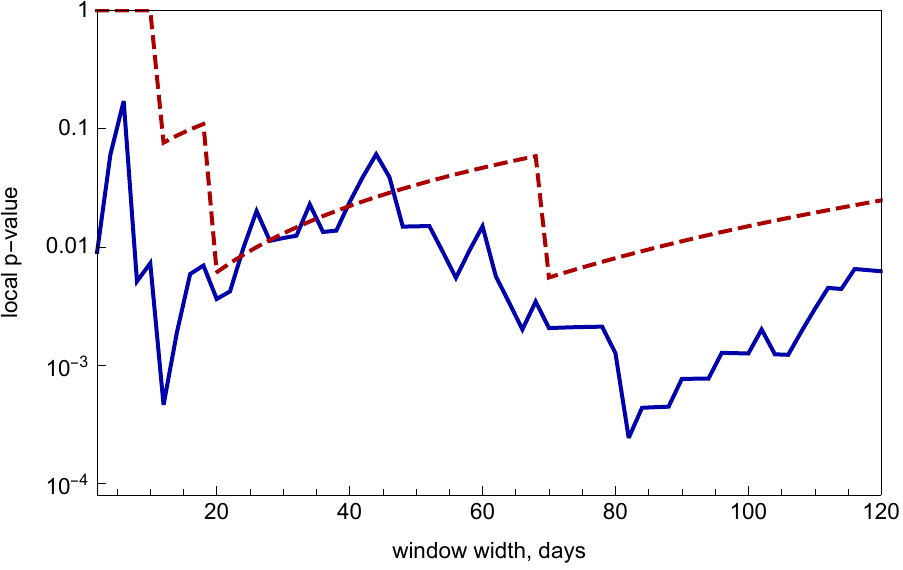}
    \caption{Dependence of the p-value on the width of the window centered on the neutrino arrival time (full line: all events, dashed line: photon median selection). See the text for details.
        }
    \label{fig:p-from-win}
\end{figure}

\begin{figure}
    \centering
    \includegraphics[width=\linewidth]{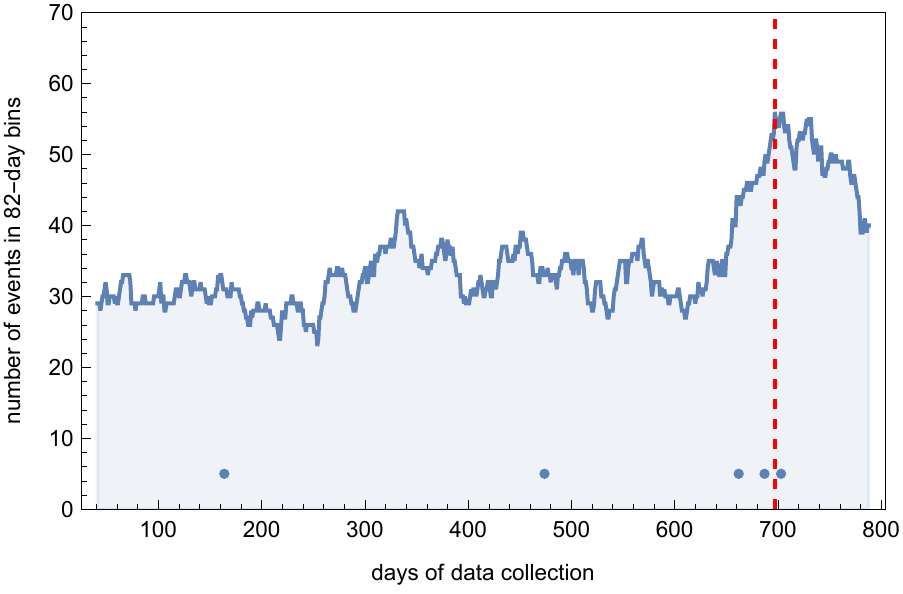}\\
  \caption{Number of events in $d=82$-day bins centered at a given day of data \red collection. Dots indicate the days of arrival of photon median candidate events\black.
  The vertical dashed line indicates the neutrino arrival time. 
        }
    \label{fig:bins82days}
\end{figure}

There are $N=346$ events with arrival directions from CCC during 829 days of \red data collection\black, of which 5 are ``photon median candidates''. Figure \ref{fig:p-from-win}
presents the probability $p(d)$ for these two sets. For the full set, $p_{\rm pre}=2.4\times 10^{-4}$ and is achieved for $d=82$ (we note however that $p(d=12)$ is almost that low). Stated in Gaussian terms, this value of $p_{\rm pre}$ would correspond to the 3.67$\sigma$ (pre-trial) significance of the flare at the neutrino arrival time. However, the correction for window-width trials reduces the significance to $p=1.5 \times 10^{-3}$ (3.17$\sigma$ post-trial). The results for photon median candidate selection alone are $p_{\rm pre}=5.5 \times 10^{-3}$ ($2.78\sigma$ pre-trial), optimal $d=70$ and $p=1.1 \times 10^{-2}$ ($2.55\sigma$ post-trial). Figure~\ref{fig:bins82days}
presents the number of events in a sliding window of the width $d=82$ days \red of data collection \black centered at a certain date, as a function of this date. One can clearly see the enhancement around the neutrino event, consistent between all events and photon median candidates. 

While the strongest signal was found for a flare with the duration of 82 days \red of data collection \black (85 calendar days), this particular number may be altered by fluctuations. The number of excess events is obtained from the time window which is tuned to have the strongest signal, so it may also be biased. The time-window correction which eliminates the effect of the flare duration tuning is applied, and the photon-flare parameters are estimated by Monte-Carlo simulations\red, see Appendix \ref{sec:appendix}\black. We obtain the flare duration of $89.5^{+32.0}_{-18.6}$~days and the source flux during this flare of $I_\gamma(E_\gamma>300~\mbox{TeV})=(5.6\pm1.8) \times 10^{-12}$~cm$^{-2}$s$^{-1}$. We also define the fluence as the flux time-integrated over the flare, $13\pm4$~GeV/cm$^2$.

It is instructive to compare this fluence in $E_\gamma>300$~TeV photons with an estimate of the fluence of the putative neutrino flare. IceCube did not find a statistically significant excess of low-energy neutrinos associated with the alert on a day scale \citep{IceCube-GCN.28946-extranu}, nor additional high-energy neutrino alerts from this direction were reported, so the rough estimate of the neutrino fluence is determined by the single alert event \citep{IceCube-GCN.28927} and, given the IceCube effective area for the BRONZE alert selection \citep{IceCube-Gold-Bronze-alerts}, is of order $\sim 3.5$~GeV/cm$^2$ \citep{CarpetATel-obs}\footnote{This gives an order-of-magnitude estimate only because of the Eddington bias in the flux estimation of a single event, cf.\ \citet{Eddington-Franckowiak}, large uncertainties in the energy estimation of track events and the lack of information about neutrino events in the days around the alert.}. Therefore, the orders of magnitude of the observed fluences are consistent with each other: in the standard pi-meson neutrino production mechanism, the energy in gamma rays is about twice the energy in neutrinos.

Small statistics and large background make it unfeasible to derive the observed gamma-ray spectrum above 300~TeV. However, we note that a comparison of the numbers of excess of events in the full data set and among the photon median candidates speaks in favor of a hard spectrum: the excess in the former is larger than twice the excess in the latter. For instance, for a very hard $E_\gamma^{-1.4}$ spectrum, only 1/3 of photons pass the median cut designed assuming $E_\gamma^{-2}$ spectrum we use. Such hard spectra do not look implausible in view of recent theoretical \citep{Bykov-hard-spectrum} and observational \citep{Dzhatdoev-hard-spectrum} results.

\begin{figure}
    \centering
    \includegraphics[width=\linewidth]{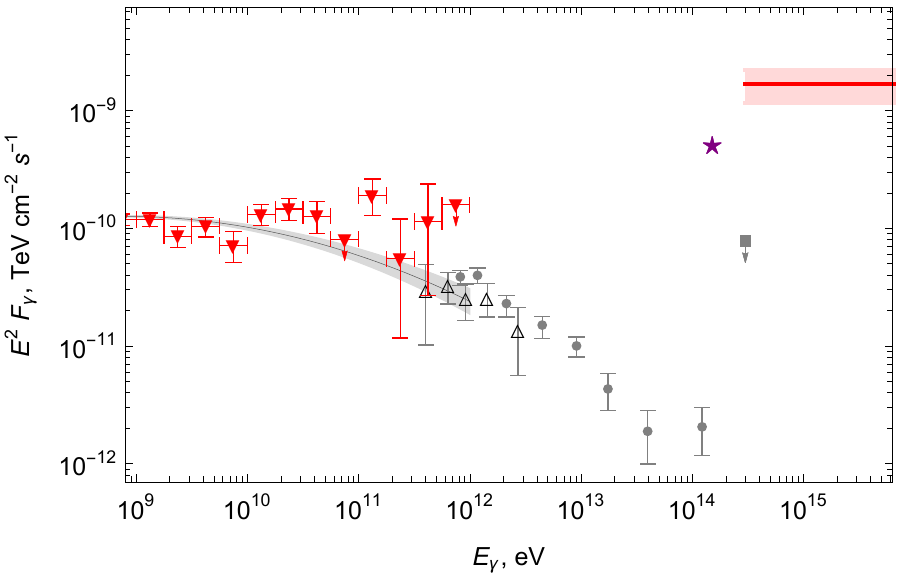}
  \caption{Spectral energy distribution of Cygnus Cocoon above 1~GeV. Gray color: time-averaged (line, 4FGL flux model \citep{4FGL}; open triangles, ARGO \citep{ARGO-Cocoon}; circles, HAWC \citep{HAWC-Cocoon}; box, Carpet--2, this work). 
  Red color: flare (triangles, Fermi LAT; line, Carpet--2 -- this work). Purple star: estimate of the neutrino fluence. See the text for details.
        }
    \label{fig:SED}
\end{figure}

 Additionally, we reconstructed the spectral energy distribution (SED $ \equiv E^2dN/dE \equiv E^2F_{\gamma}$) of Cygnus Cocoon in the energy range from $100$~MeV to $1$~TeV averaged over the same $d=82$-day period around the neutrino arrival
 using publicly available data of the Fermi Large Area Telescope (Fermi-LAT) \citep{Fermi-LAT}.
 The region of interest (ROI) in our analysis was a $15^{\circ}$ square centered at the 4FGL position of the Cygnus Cocoon. Making use of \textit{fermitools}\footnote{\url{https://github.com/fermi-lat/Fermitools-conda}} (version 2.0.8) and \textit{fermipy} (version 1.0.1) \citep{Fermipy} packages and the instrument response function \texttt{P8R3\_SOURCE\_V3}, we constructed a model of the observed gamma-ray emission from the ROI containing 4FGL sources, including our source of interest \object{4FGL~J2028.6$+$4110e}; models of the isotropic gamma-ray background \texttt{iso\_P8R3\_SOURCE\_V3\_v1} and the galactic diffuse emission \texttt{gll\_iem\_v07}. Normalizations of both diffuse backgrounds and the spectral shape of the galactic diffuse emission, along with all spectral parameters of the source of interest, were left free. The normalizations for other sources within $5^{\circ}$ from the ROI center were also left free, but their spectral shapes and all parameters of sources beyond the $5^{\circ}$-circle were fixed to the catalog values. Other event selection parameters were set according to the standard recommendations of the Fermi-LAT collaboration for a galactic point source analysis\footnote{\url{https://fermi.gsfc.nasa.gov/ssc/data/analysis/documentation/Cicerone/Cicerone_Data_Exploration/Data_preparation.html}}.
 Using \texttt{GTAnalysis.sed} \textit{fermipy} method we obtained the SED of \object{4FGL~J2028.6$+$4110e} averaged over the $82$-day flare period. We found an indication ($\sim2.17\sigma$ significance) of the spectrum hardening with respect to the spectrum of the Cygnus Cocoon presented in the 4FGL catalog.
Figure~\ref{fig:SED}
compares our observations with other high-energy data. \red Unfortunately, no simultaneous \black
data on the month-scale flare related to the neutrino arrival have yet been  published by other experiments. 

\red The source entered the field of view of Carpet--2 16 minutes after the \black neutrino arrival. Like HAWC \citep{HAWC-GCN.28952}, we do not find a significant flux enhancement within 24~hours from \red the neutrino \black \citep{Carpet-Atel-1day}. However, we observed a very unusual cluster of events at the scale of minutes on the day of the neutrino alert; its significance and implications will be discussed elsewhere. 
\section{Conclusions}
\label{sec:concl}
An excess of events was observed by Carpet--2 from the direction of the Cygnus \red region \black in temporal coincidence with the IceCube neutrino alert from the same direction. Statistical significance of the excess is 3.1$\sigma$ post-trial. The excess may be interpreted as a $E_\gamma>300$~TeV photon flare with the duration of $\sim3$ months around the neutrino event and the fluence of $13\pm 4$~GeV/cm$^2$. For the first time, rare sub-PeV neutrino and gamma rays from the direction of a prospective Galactic PeVatron were observed in directional and temporal coincidence. This observation supports previously proposed \red scenarios \black of the origin of a part of observed high-energy neutrinos in pi-meson decays in Galactic sources. 
\red Note that poor localization of the neutrino event, as well as modest angular resolution of Carpet-2, leave open the possibility of the association of these events with other interesting sources in CCC, including the gamma-ray loud microquasar Cyg~X-3, gamma-ray binary PSR~J2032$+$4027 etc. \black

This possible sub-PeV flare may be searched in the recorded data of other gamma-ray air-shower experiments, LHAASO \citep{LHAASO}, HAWC \citep{HAWC}, Tibet \citep{TIBET}, GRAPES-3 \citep{GRAPES-3} and TAIGA \citep{TAIGA}, as well as of neutrino telescopes, IceCube \citep{IceCube} in the track mode and ANTARES \citep{ANTARES} and Baikal-GVD \citep{Baikal-GVD} in the cascade mode. 
Future monitoring of the source by these instruments, as well as by the upgraded Carpet--3 \citep{Carpet-3}, is also encouraged.
\acknowledgments
We thank A.~Bykov\red, T.-Q.~Huang, K.~Kawata \black and D.~Semikoz for illuminating discussions.
This work is supported by the Ministry of science and higher education of Russian Federation under the contract 075-15-2020-778. 
E. I. P. thanks the Foundation for the Advancement of Theoretical Physics and Mathematics ``BASIS'' (Contract No. 20-2-10-7-1) and the Non-profit Foundation for the Development of Science and Education ``Intellect'' for the  student scholarships.
%
\facilities{Carpet-2, IceCube neutrino observatory, Fermi-LAT}




\red
\appendix

\section{Estimation of the flare duration and fluence}
\label{sec:appendix}
\paragraph{Estimation of the number of signal events.} Assume that the flare flux corresponds to $x$ photons during the flare period and the photon spectrum is $E^{-2}$. Of them, on average, we expect the excess of $x/2$ photon median candidates and $x/2$ other events above the background rates of $b_1\approx 0.25$ for photon median candidates and $b_2\approx 32.2$ for the rest of events (determined from the off-flare period of observations). We thus expect to observe $b_1+x/2$ photon candidates and $b_2+x/2$ other events during the flare, but all these numbers fluctuate. To determine $x$, we maximize the probability to observe the actual numbers of events, $o_1=3$ photon candidates and $o_2=53$ other events, simultaneously. We find $x\approx 9.9 \pm 3.2$; this number is divided by the effective exposure to determine the flux. Note that the probability to observe $\le o_1$ and $\ge o_2$ events in the respective data sets is $\approx 0.07$ for this $x$, so the two observations are consistent at the 93\% CL. 

\paragraph{Correction of the biases caused by trials.} The flare window was chosen such that the excess is most significant. This procedure selects a positive fluctuation in the number of observed, signal plus background, events (the post-trial significance accounts for this). Thus this best duration of the flare and the corresponding excess flux are biased. To correct for this effect, we perform a Monte-Carlo simulation of the entire procedure, assuming the flare parameters determined above. For each simulated realization of the events, we find the strongest-signal flux and duration. We then compare these reconstructed fluxes and durations with those assumed in the simulation. The reconstructed flux and duration differ from the true values by factors of 0.86 and 1.08, respectively. These coefficients are accounted for in the values reported in the main text.

\black
\bibliography{Cygnus}{}
\bibliographystyle{aasjournal}




\end{document}